\documentstyle[pramana,epsf,floats]{ias}
\begin{document}
\mark{{Excitation of Pygmy Dipole Resonance}{A Vitturi et al.}}
\title{Excitation of Pygmy Dipole Resonance in neutron-rich nuclei via Coulomb and nuclear fields}

\author{A Vitturi $^{(1)}$, EG Lanza $^{(2)}$, MV Andr\'es $^{(3)}$, F
  Catara $^{(2)}$ and D Gambacurta $^{(2)}$}
\address{(1) Dipartimento di Fisica  and INFN, Padova, Italy \\
  (2) Dipartimento di Fisica and INFN, Catania, Italy \\
  (3) Departamento de FAMN, Facultad de F\'{i}sica, Sevilla, Spain }
\keywords{Exotic nuclei, pygmy dipole resonance} 
\pacs{21.60.Ev, 21.60.Jz, 25.60.-t, 25.70.-z} 
\abstract{ We study the nature of the
  low-lying dipole strength in neutron-rich nuclei, often associated
  to the Pygmy Dipole Resonance.  The states are described within the
  Hartree-Fock plus RPA formalism, using different parametrizations of
  the Skyrme interaction.  We show how the information from combined
  reactions processes involving the Coulomb and different mixtures of
  isoscalar and isovector nuclear interactions can provide a clue to
  reveal the characteristic features of these states.  }

\maketitle
\section{Introduction}
Large interest has been devoted in the last few years to the evolution
of the properties of collective states in neutron-rich nuclei.
Besides the usual giant multipole resonances (as the Giant Dipole
State, GDR), special attention has been focussed on the presence of
dipole strength at low excitation energy, of the order of few per cent
of the corresponding Energy Weighted Sum Rule (EWSR)\cite{noi}. This
strength has been often associated to the possible existence of a
collective mode of new nature, corresponding to the oscillation of the
valence neutron skin against the proton plus neutron core (Pygmy
Dipole Resonance, PDR) (see, for example ref.~\cite{colo} and
reference quoted in).

From a theoretical point of view the presence of this low-lying
strength is predicted by almost all microscopic models, ranging from
Hartree-Fock plus RPA with Skyrme interactions to relativistic
Hartree-Bogoliubov plus relativistic quasiparticle RPA.  All these
approaches predict similar amounts of strength, but often disagree on
the collective (or not) nature of these states, on their fragmentation
and on their isoscalar/isovector contents\cite{colo}.

From an experimental point of view the pulling evidence for these
states has come from high-energy Coulomb excitation
processes\cite{GSI}.  As known, these can only provide values of the
multipole B(E$\lambda$) transition rates, being instead unable to
provide further information on wave functions and transition
densities, which are the necessary ingredients to characterize the
nature of these states.  A part of this information can be obtained by
resorting to reactions where the nuclear part of the interaction is
involved.  By tuning the projectile mass, charge, bombarding energy
and scattering angle one can alter the relative role of the nuclear
and Coulomb components, as well as of the isoscalar and isovector
contributions.

In this talk we present the predictions for the excitation of the
low-lying (PDR) and high-lying (GDR) dipole states in the neutron-rich
$^{132}$Sn by different projectiles ($\alpha$, $^{40}$Ca, $^{48}$Ca)
at different bombarding energies.  The dipole states, their wave
functions and the corresponding transition densities have been
obtained within the Hartree-Fock plus discrete RPA with Skyrme
interactions.  Formfactors have then been obtained by double-folding
the M3Y N-N interactions, and then used to describe the quantal
evolution of the system along the classical ion-ion trajectories. We
will show how the excitation probabilities are sensitive to the
details of the transition densities (and not simply to the B(E1)
values) and how these can be probed by combination of different
processes.
\begin{figure}[htbp]
\epsfxsize=13cm
\centerline{\epsfbox{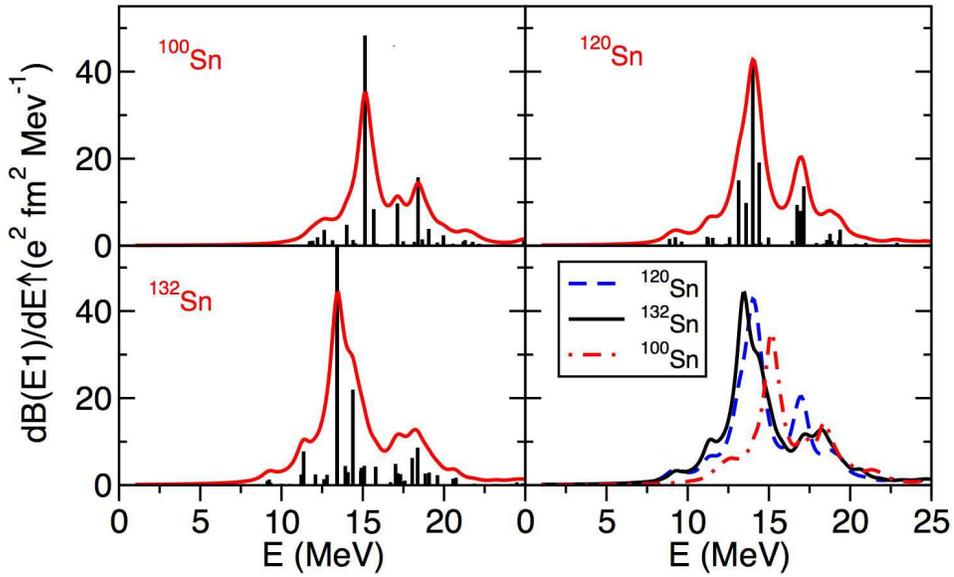}}
\caption{Isovector strength distributions for dipole states for tin
  isotopes calculated with the SGII interaction. The bars are the RPA
  B(E1) values. The solid curves represent $dB(E1)/dE$ as obtained by
  adopting a smoothing procedure described in the text. In the lower
  right figure we report the same continuous lines shown in the other
  frames. }
\label{fig:1}
\end{figure}

\section{RPA dipole strength distribution and Coulomb excitation probabilities}

Predictions for the dipole response in many-body systems are provided
by different approaches.  At the level of particle-hole plus
ground-state correlations, a notable role has always been played by
theories based on the Random Phase Approximation, both in their
non-relativistic and relativistic versions.  We will use here the
results obtained in the simplest discrete non-relativistic RPA
approach with Skyrme interactions, consistently used both at the level
of mean-field Hartree-Fock and of RPA.  An example of the evolution of
the dipole response with the neutron number is shown in
Fig.~\ref{fig:1}, where the three isotopes $^{100,120,132}$Sn are
separately considered (from Ref.\cite{lan}).  Note that the bars are
the discrete response of the RPA calculation while the continuous
lines are generated by a smoothing procedure using a Lorentzian with a
1 MeV width. The continuous lines are drawn only to easily see where
the major strength is located.  To better appreciate the isotope
dependence, the three distributions of dipole strength are shown
together in the lower (right) frame in Fig.~\ref{fig:1}.  As one can
see, together with the usual lowering of the energies of the dominant
Giant Dipole Resonance with the increased mass number, we have the
appearance of some low-lying strength (carrying a fraction of the EWSR
of the order of few per cent) below 10 MeV. These are precisely the
states that are candidates to be interpreted as Pygmy Dipole
Resonances, associated with the occurrence of neutron skins in the
nuclear densities and their oscillations with respect to the
proton+neutron cores.

\begin{figure}[htbp]
\epsfxsize=11cm
\centerline{\epsfbox{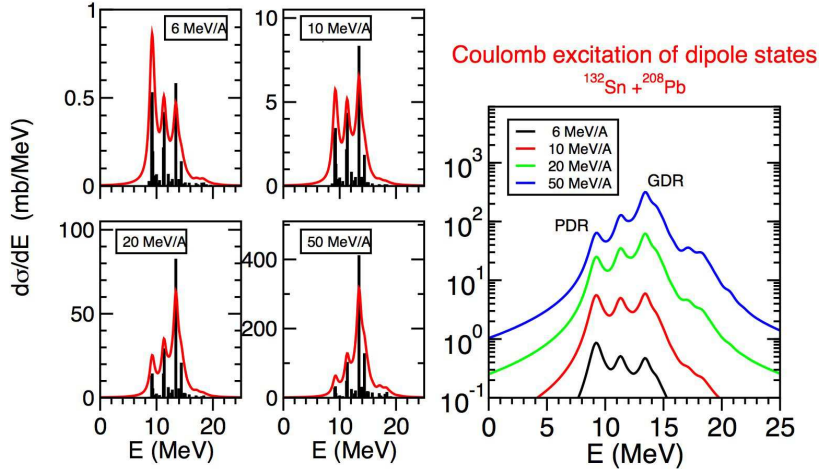}}
\caption{Coulomb excitation cross section for the system $^{132}$Sn
  +$^{208}$Pb at several incident energies as indicated in each
  frame. The bars correspond to the calculations done for each of the
  RPA states while the continuous lines are obtained with the
  smoothing procedure described in the text. In the frame on the right
  we put together the solid lines in order to show the relative
  evolution of the states as function of the incident energies.  }
\label{fig:2}
\end{figure}

The dipole strength distribution can be directly tested in Coulomb
excitation processes.  Calculated Q-value distributions $d\sigma_C/dE$
of total Coulomb excitation cross sections for the $^{132}$Sn +
$^{208}$Pb reaction at different bombarding energies are shown in
Fig.~\ref{fig:2}.  As expected, at high bombarding energies the cross
sections just follow the B(E1) strength distribution but at lower
energy the kinematical cut-off enhances the role of the states with
lower energies.  As apparent from the lower frame (right) where the
different cross sections are combined, at very low bombarding energy
the probability of exciting the ``PDR" region becomes even larger than
exciting the GDR.

\section{Isoscalar and isovector transition densities and excitation
  of dipole states via nuclear fields}

\begin{figure}[htbp]
\epsfxsize=12cm
\centerline{\epsfbox{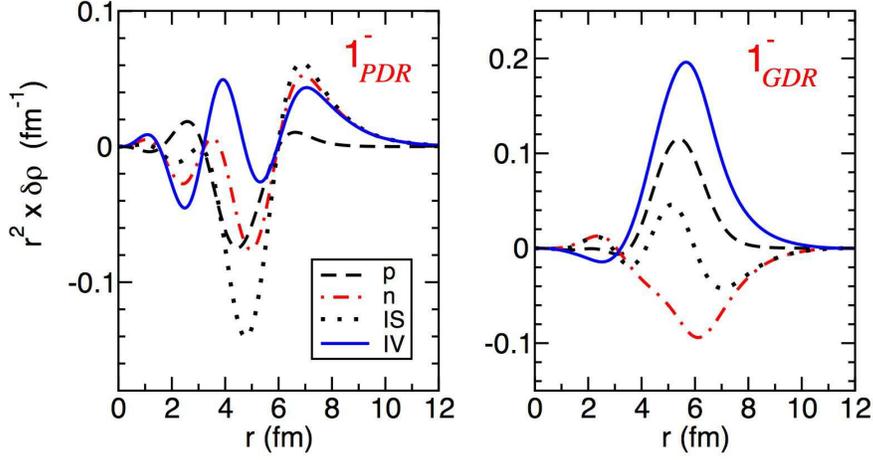}}
\caption{Transition densities for the low-lying dipole state (PDR)
  (left) and for the GDR (right) for the $^{132}$Sn isotope calculated
  with the SLY4 interaction. We show the proton, neutron, isoscalar
  and isovector components (as indicated in the legend).}
\label{fig:3}
\end{figure}

\begin{figure}[htbp]
\epsfxsize=12cm
\centerline{\epsfbox{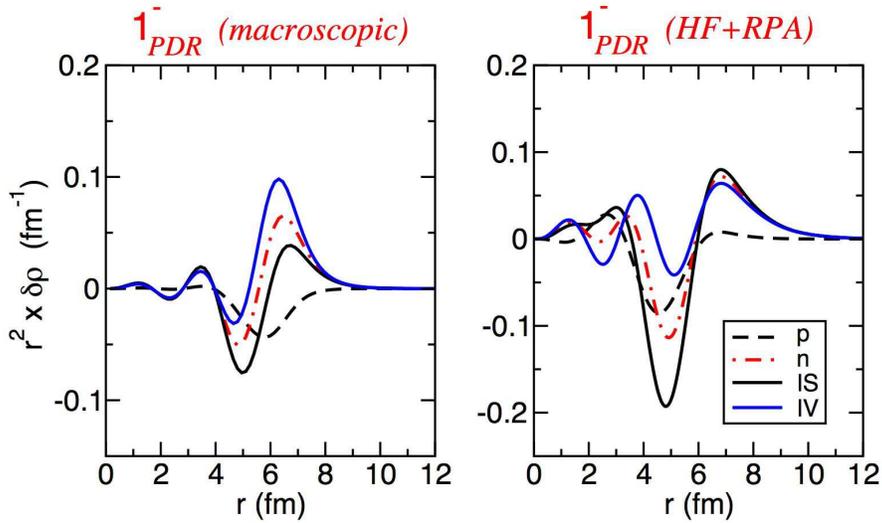}}
\caption{Transition densities for the low-lying dipole state for the
  $^{132}$Sn isotope. The frame on the left show the macroscopic
  transition densities according to eq.~(\ref{eq:td}), the ones on the
  right are calculated microscopically with the HF + RPA. We show the
  proton, neutron, isoscalar and isovector components (as indicated in
  the legend).}
\label{fig:4}
\end{figure}

Excitation processes via the Coulomb field can only test the
$B(E\lambda)$-values, i.e. integrated electromagnetic matrix elements.
More precise information on the specific nature of the states is
instead embedded in the corresponding transition densities.  As an
example we show in Fig.~\ref{fig:3} the RPA transition densities
associated with the GDR (right frame) and with a state in the ``PDR"
region (left frame) in $^{132}$Sn.  Neutron and proton components of
the transition densities are separately shown, together with their
isoscalar and isovector combinations. The two cases clearly display
very different behaviours.  The one associated with the GDR shows the
usual opposite-phase behaviour of the proton and neutron components,
leading to a dominant isovector character.  As known \cite{noi2},
however, in these very-neutron rich nuclei the presence of different
radii for the proton and neutron densities leads to non-vanishing
isoscalar transition densities, opening the possibility of exciting
the GDR also via isoscalar probes.
The situation is rather different in the case of the other state at
lower energy.  Here neutron and proton components seem to oscillate in
phase in the interior region, while having a pure neutron oscillation
(with opposite phase) in the surface region. This may score a point in
favor of the interpretation of this state as a Pygmy Dipole Resonance,
macroscopically described as the oscillation of the neutron skin with
respect to the proton+neutron cores.  Such behaviour, which has been
found also in all the other microscopic approaches\cite{colo,tsole},
can be taken as a sort of definition of PDR.  A macroscopic
description of such a mode assumes a separation of the neutron density
into a core part $\rho_N^C$ with $N_C$ neutrons and a valence part
$\rho_N^V$ with $N_V$ neutrons ($N=N_C+N_V$), with a proton density
$\rho_P$ with $Z$ protons.  This leads to neutron and proton
transition densities given by
\begin{eqnarray}
{\delta \rho_N(r)} &=& \beta \left[ {N_V \over A}~{d\rho_N^C(r) \over dr}-{N_C+Z \over A}
{d\rho_N^V(r) \over dr} \right]
\nonumber \\
{\delta \rho_P(r)} &=& \beta \left[ {N_V \over A}~{d\rho_P(r) \over dr}  \right]
\label{eq:td}
\end{eqnarray}
\noindent
with $\beta$ a proper strength parameter.  The microscopic RPA and the
macroscopic transition densities, normalized to the same B(E1) value,
are compared in Fig.\ref{fig:4}.
Although some similarities are present, a full interpretation of the
state in terms of the PGR is not obvious.  It should be noted that,
besides the requirement of the shape of the transition density, the
macroscopic picture should also involve a collective nature of the
state, therefore implying some coherent contribution of the different
particle-hole in the RPA state, This requirement was not found to be
fulfilled at least in the calculations of ref.\cite{lan}.

The transition densities are the basic ingredients to construct the
nuclear formfactors describing nuclear excitation processes.  These
formfactors can be obtained by doublefolding \cite{sat} the transition
densities with the density of the reaction partner and the $N-N$
interaction (taken in our case as M3Y), including both isoscalar and
isovector terms.  The relative role of isoscalar and isovector
formfactors depends on the corresponding mixture in the transition
density of the specific state but also on the nature of the specific
reaction partner (i.e purely isoscalar in the case of
$\alpha$-particle or other $N=Z$ nuclei, isoscalar+isovector for $N
\neq Z$ nuclei).  The change of reaction (and of the bombarding
energy), with the consequent change of the relative role of nuclear
and Coulomb components as well as of the isoscalar/isovector nuclear
components) will alter the relative population of the different
states.  Viceversa, although more tricky, from this different
population one can hope to extract the different properties of the
transition densities, and hence the different nature of the states.

\begin{figure}[htbp]
\epsfxsize=12cm
\centerline{\epsfbox{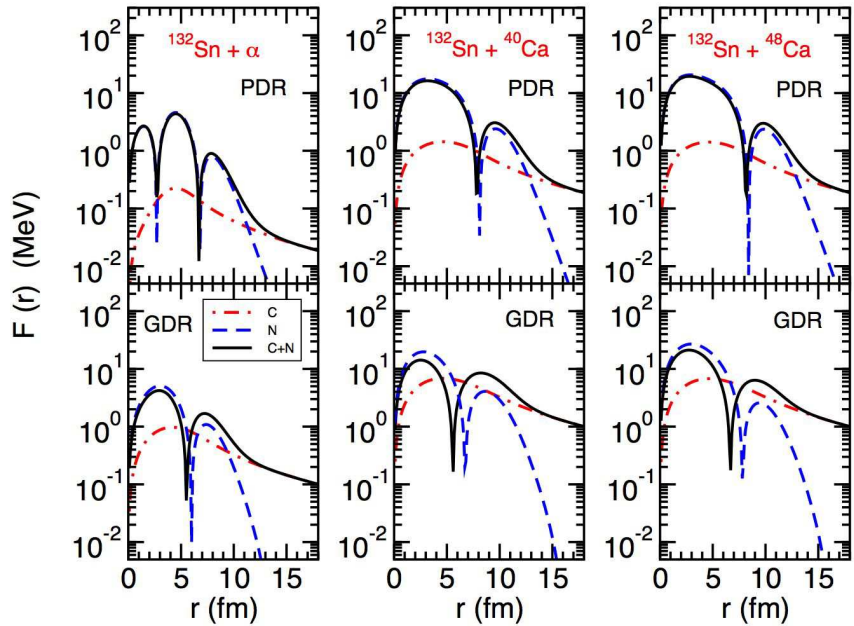}}
\caption{Formfactors for three different systems $^{132}$Sn +
  $\alpha$, $^{40}$Ca, $^{48}$Ca. The upper parts refer to the PDR
  states while the lower ones are for the GDR. The different component
  are shown together with the total one (solid black line). }
\label{fig:5}
\end{figure}

We will consider here the excitation of the Pygmy and Giant dipole
states in $^{132}$Sn by different partners: $\alpha$, $^{40}$Ca and
$^{48}$Ca.  The corresponding formfactors (nuclear, Coulomb and total)
and shown in Fig.~\ref{fig:5}.

\begin{figure}[htbp]
\epsfxsize=12cm
\centerline{\epsfbox{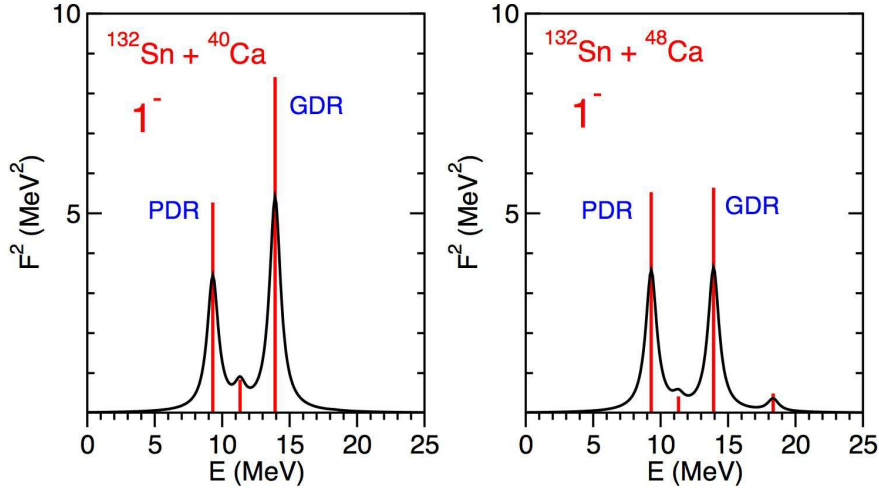}}
\caption{Square of the formfactor for the systems indicated in the
  figure and for four significative dipole states. The continuous
  lines are obtained with the same smoothing procedure used
  previously.}
\label{fig:6}
\end{figure}

For an easier comparison, we compare in Fig.~\ref{fig:6} the square of
the formfactors at the surface for the excitation of different states.
We have chosen, for simplicity, four significative dipole states which
correspond to the four peaks present in the B(E1) strength
distributions for $^{132}$Sn of Fig.~\ref{fig:1}. One can see how the
different reactions alter the relative ``intensities" of PRD and GDR
states. This is due to the different interplay of isoscalar and
isovector contributions.

\begin{figure}[htbp]
\epsfxsize=8cm
\centerline{\epsfbox{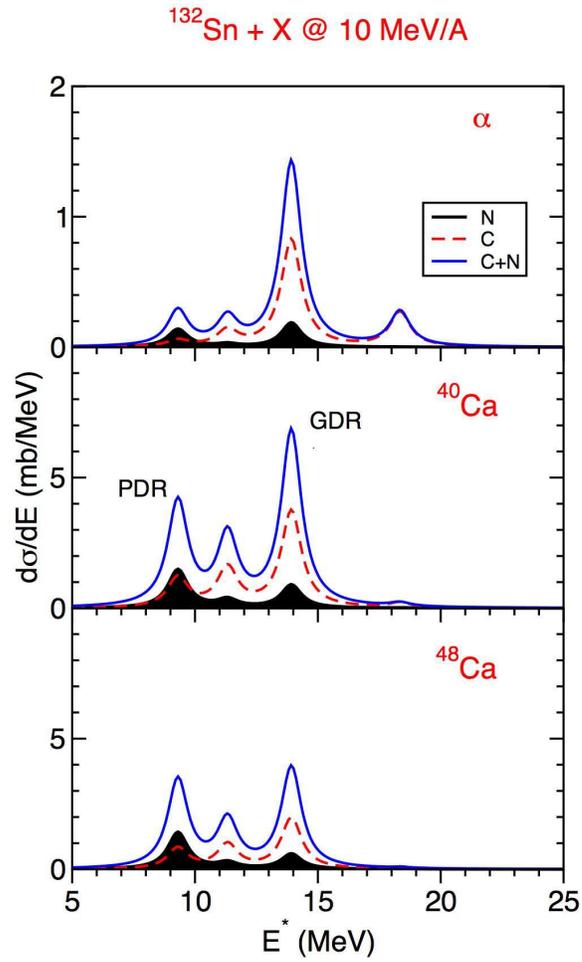}}
\caption{Differential cross sections as function of the excitation
  energy for the systems $^{132}$Sn + $\alpha$, $^{40}$Ca, $^{48}$Ca
  at 10 MeV per nucleon.  The Coulomb contribution is shown as (red)
  dashed line. The shaded (black) area corresponds to the nuclear
  contribution.}
\label{fig:7}
\end{figure}
\begin{figure}[htbp]
\epsfxsize=12cm
\centerline{\epsfbox{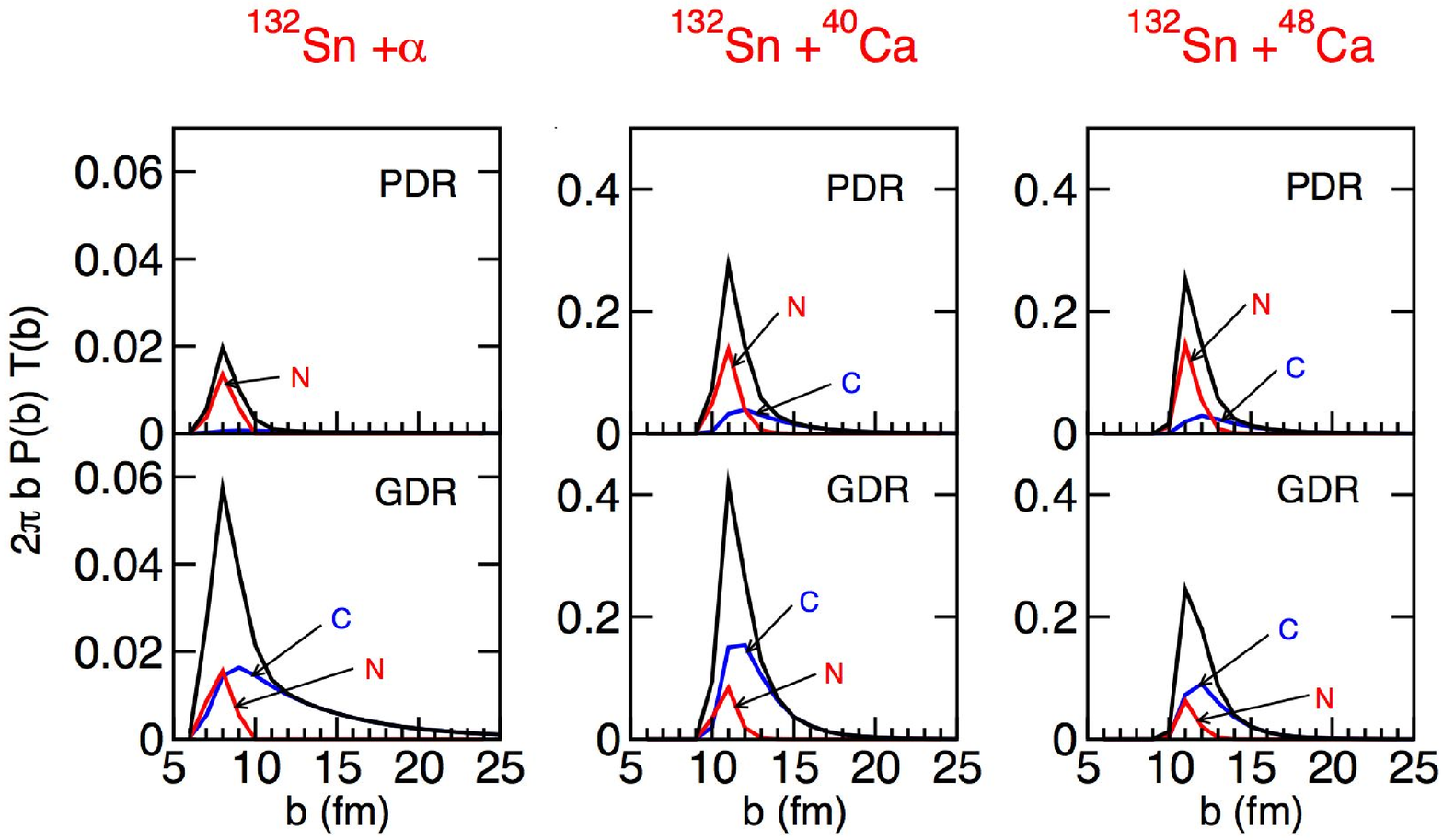}}
\caption{Partial wave cross sections as function of the impact
  parameter b for the systems $^{132}$Sn + $\alpha$, $^{40}$Ca,
  $^{48}$Ca at 10 MeV per nucleon.  The Coulomb (C) and nuclear (N)
  contributions are indicated in the figures. The curves with the
  highest maximum are the total cross section. }
\label{fig:8}
\end{figure}

With ion-ion potential and formfactors we can now calculate cross
sections (for example within the semiclassical approach as done in
ref.~\cite{lan}).  The energy differential total cross sections are
shown in Fig.~\ref{fig:7}.  The different contributions from Coulomb
and nuclear formfactors are separately show.  It is clear that the
balance between PDR and GDR vary in the different reactions.  The
ratios can be further modified by looking at the differential angular
distributions.  In the semiclassical picture, these are associated to
different ranges of impact parameters.  Nuclear contributions are
known to be enhanced at grazing angles, corresponding to grazing
impact parameters.  This is clearly evidenced in Fig.~\ref{fig:8},
where the partial-wave cross sections are shown as a function of the
impact parameter.

\section{Conclusions}
The interpretation of the low-lying dipole strength in very-neutron
rich nuclei as a Pygmy dipole state of collective nature needs to be
carefully checked. Valuable information on the nature of these states
can be obtained by excitation processes involving the nuclear part of
the interaction, which are sensitive to the shape of the transition
densities. The use of different bombarding energies, of different
combinations of colliding nuclei involving different mixture of
isoscalar/isovector components, together with the mandatory use of
microscopically constructed formfactors, can provide the clue towards
the solution of the problem.

\end{document}